\documentclass[twocolumn,superscriptaddress]{revtex4-2}
\usepackage{amssymb}
\usepackage{amsmath}
\usepackage{graphicx}
\usepackage{color}

\begin{document}

\title{Quasi-lattices of qubits for generating inequivalent multipartite entanglements}

\author{Hou Ian}

\affiliation{Institute of Applied Physics and Materials Engineering, FST, University of Macau, Macau}

\begin{abstract}
The mesoscopic scale of superconducting qubits makes their inter-spacings
comparable to the scale of wavelength of a circuit cavity field to
which they commonly couple. This comparability results in inhomogeneous
coupling strengthes for each qubit and hence asynchronous Rabi excitation
cycles among the qubits that form a quasi-lattice. We find that such
inhomogeneous coupling benefits the formation of multi-photon resonances
between the single-mode cavity field and the quasi-lattice. The multi-photon
resonances lead, in turn, to the simultaneous generation of inequivalent
$\left|\mathrm{GHZ}\right\rangle $ and \emph{$\left|W\right\rangle $}
types of multipartite entanglement states, which are not transformable
to each other through local operations with classical communications.
Applying the model on the 3-qubit quasi-lattice and using the entanglement
measures of both concurrence and 3-tangle, we verify that the inhomogeneous
coupling specifically promotes the generation of the totally inseparable
$\left|\mathrm{GHZ}\right\rangle $ state.
\end{abstract}

\maketitle

\section{Introduction}

Superconducting qubits are artificial two-level atoms made from superconducting
islands linked by multiple Josephson junctions~\cite{clarke08}.
When coupled to a microwave field via a transmission line or coplanar
waveguide resonator in the circuits, they become a circuit quantum
electrodynamic (QED) system~\cite{blais04}. Circuit QED is equivalent
to cavity QED for natural atoms and circuit versions of optical effects~\cite{jqyou11},
such as lasing~\cite{ashhab09,andre09}, electromagnetically induced
transparency~\cite{ian10,sanders10,sun14}, and parametric amplification~\cite{lehnert07},
have already been demonstrated.

Over the years, the studies on circuit QED have been extended to multi-qubit
systems~\cite{fink09,macha13} and their entanglement property~\cite{mlynek12},
superradiant states~\cite{filipp11,mlynek14} as well as phase transitions~\cite{feng15}
have been experimentally explored. In the fabricated circuits, these
artificial atoms are orders of magnitude (usually $10^{3}$ to $10^{4}$
in linear scale) larger than their natural counterparts. To make existing
theories and models for optical cavities relevant to circuit systems,
the experimental remedies are: (i) placing the qubits exactly at the
antinodes of the resonator mode~\cite{fink09} so that Tavis-Cummings
model becomes applicable; (ii) by stretching the transmission line
such that the qubits are congregated at the center~\cite{macha13}
so that the Dicke model adopted in optomechanical system~\cite{colombe07,ian08,hunger10}
becomes applicable; or (iii) by fabricating dedicated resonators for
each qubit and externally synchronizing the drivings to each resonator~\cite{lucero12}.
All of these approaches effectively reduce the multi-qubit systems
back to conventional scenarios one meets in optical cavities. However,
under a general scenario, the inter-spacing between two neighborings
qubits is comparable with the resonator mode wavelength and the qubits
do not necessarily lie on the anti-nodes, letting the qubit-resonator
coupling become inhomogeneous and rendering conventional theoretical
models irrelevant for multi-qubit circuits.

Here, we consider a general multi-qubit system as a quasi-lattice:
a lattice that has each artificial atom located at a fixed lattice
point and two neighboring artificial atoms distanced by a fixed lattice
spacing, though each artificial atom receives a waveform-dependent
coupling strength to a single-mode cavity field. That means, we disregard
the Dicke model that only considers the scaling effect of coupling
and consider rather the Rabi frequency of each qubit to be uneven
such that the excitations throughout the quasi-lattice are asynchronous.
Based on the projected deformed SU(2) algebra method~\cite{ian12},
we theorize the quasi-lattice as a large collective spin that regard
the lattice-resonator interaction as a quasi-particle excitation.
The quasi-particle states are a set of multi-photon stationary resonant
states that diagonalize the interaction Hamiltonian. Then, by employing
the entanglement measures of concurrence~\cite{wootters98,carvalho04}
and 3-tangle~\cite{coffman00} on these states, we determine that
both the $\left|\mathrm{GHZ}\right\rangle $ and the $\left|W\right\rangle $
entangled states are cogenerated simultaneously during the resonant
processes. In particular, with the quasi-lattice being asynchronously
excited, the generation of $\left|\mathrm{GHZ}\right\rangle $ state
is optimized over past homogeneous-coupling schemes~\cite{wei06,matsuo07}.
By linking the asynchronicity statically to a deformation factor of
the underlying algebra, one can associate maximized entanglement measures
that characterizes the $\left|\mathrm{GHZ}\right\rangle $ state at
specific scenarios of asynchronous excitation. The entanglement optimization
was proved earlier to be controllable via the coupling strength when
two qubit are directly capacitatively coupled to each other~\cite{zhang09}.

Since the $W$ and the GHZ states are inequivalent to each other~\cite{dur00,bennett00},
i.e. one cannot obtain one state from the other through local operations
with classical communications (LOCC), the ability to generate both
in a unified system greatly reduces our reliance on complex quantum
gate routines currently employed~\cite{lucero12}. Especially, the
difficulty in generating GHZ-type entanglement shows the potential
of the quasi-lattice structure as a fundamental building block of
future quantum information processing devices. This peculiar feature
of the quasi-lattice can be attributed, following Dicke's notions,
to its avoidance of full population inversion because of the asynchronous
excitations and thus yielding a higher cooperation among the qubits.
Consequently, the quasi-lattice receives higher probability in absorbing
multiple photons from the resonator, resulting in a more entangled
state than it would be achievable with homogeneous coupling. Moreover,
it was shown that the mechanism to generate \emph{W}-states becomes
faster when the coupling is inhomogeneous~\cite{lopez12}, which
verifies from another angle that the inhomogeneous coupling of quasi-lattices
facilitates entanglement generation.

\begin{figure}
\includegraphics[clip,width=8.2cm]{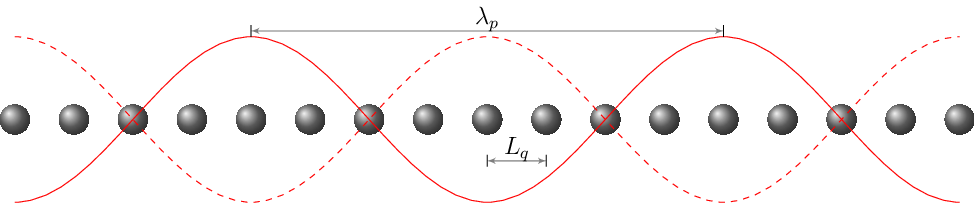}
\raggedright{}\caption{(color online) Illustrations of inhomogeneous interactions on a quasi-lattice.
Uniform lattice spacing $L_{\mathrm{q}}$ between two neighboring
qubits is on the same order of magnitude as the wavelength $\lambda_{\mathrm{p}}$
of the transmission line cavity resonator. Their ratio $\ell$ determines
the quasi-particle excitation on the lattice. The illustrated quasi-lattice
with 17 qubits has $\ell=1/4$.~\label{fig:schematics}}
\end{figure}

\section{Quasi-lattice}

To identify the problem more clearly, we model the system on the time-independent
Hamiltonian ($\hbar=1$)
\begin{equation}
H=\omega_{0}a^{\dagger}a+\sum_{j=0}^{N-1}\left[\omega_{\mathrm{q}}\sigma_{j,z}+g_{j}(\sigma_{j,+}a+\sigma_{j,-}a^{\dagger})\right].\label{eq:ham}
\end{equation}
Under the formulation of circuit QED~\cite{blais04}, each superconducting
qubit is approximated by a two-level system whose eigenenergy $\omega_{\mathrm{q}}=\sqrt{E_{C}^{2}+E_{J}^{2}}$
is diagonalized from the tunable charge energy $E_{C}$ and junction
energy $E_{J}$ of the Josephson junctions. We denote the qubit at
the $j$-th position by the Pauli matrix $\sigma_{j,z}$ and the single-mode
transmission line resonator by the pair of creation and annihilation
operators $a^{\dagger}$ and $a$. We consider all $N$ qubits be
tuned uniform at the eigenfrequency $\omega_{\mathrm{q}}$ that is
closed resonant to the resonator frequency $\omega_{0}$. Assuming
$N$ is not sufficiently large on the mesoscopic scale to warrant
collective effects on the virtual photon processes, the interaction
between each qubit and the resonator is a Jaynes-Cummings type dipole-field
coupling, where the non-rotating terms are discarded.

The lattice spacing $L_{\mathrm{q}}$, the distance between two neighboring
qubits, is assumed uniform throughout the quasi-lattice. Since we
consider $L_{\mathrm{q}}$ to be comparable to the wavelength $\lambda_{\mathrm{p}}$
of the resonator mode, we find a sinusoidal dependence of the coupling
strength $g_{j}$ on the discrete position $j$ relative to the waveform
shape of the standing cavity mode: $g_{j}=g\cos(j\pi\ell)$. In this
inhomogeneous coupling, $\ell=2L_{\mathrm{q}}/\lambda_{\mathrm{p}}$
is the dimensionless reduced qubit spacing with respect to the wavelength
$\lambda_{\mathrm{p}}$, where the factor $2$ is added to simplify
the derivations given below.

Illustrated in Fig.~\ref{fig:schematics}, we can regard $\ell$
as the relative lattice spacing and $j$ as the dimensionless integer
coordinate in units of $L_{\mathrm{q}}$ for the one-dimensional quasi-lattice.
One can see that different sets of qubits will be driven at different
Rabi frequencies by either variably spacing the qubits or controlling
the harmonic modes in the transmission line resonator~\cite{ian14}.
If each qubit is driven at the same rate by the resonator, neighboring
qubits will become fully excited in a synchronous fashion. The spontaneous
radiation rate of each individual qubit reaches maximum at the same
rate. Consequently, the probability of simultaneous absorption of
multiple photons is low. In contrast, if the quasi-lattice has inhomogeneous
coupling to the resonator mode, neighboring quits will become fully
excited at asynchronous Rabi frequencies. That means while some qubit
is spontaneously emitting photons, others which are not fully excited
not only absorb the resonator photons but also have high probability
to reabsorb the emitted photons from neighboring qubits. This circumstance
leads to a high probability of multi-photon lattice-cavity resonance.

\section{Quasi-particles}

To verify that the inhomogeneous coupling indeed increases the probability
of multi-photon resonance in the qubits, we examine the eigenspectrum
of the collective quasi-particle excitations in the quasi-lattice.
For such collective excitations, each quantized energy level does
not correspond to the energy level of the excited state of a particular
qubit, but to a delocalized exciton with different levels of excitation
probability distributed across the quasi-lattice as a whole.

To give analytical expressions for the eigenstates of the quasi-particles,
we normalize the inhomogeneous coupling to the nominal maximum coupling
rate $g$ and incorporate the inhomogeneous factors into the ladder
operators of a large collective spin. That is, we write the interaction
part of Eq.~(\ref{eq:ham}) as $H_{\mathrm{int}}=g(S_{+}a+S_{-}a^{\dagger})$
where $S_{\pm}=\sum_{j}\cos(j\pi\ell)\sigma_{j,\pm}$. Correspondingly,
the free Hamiltonian can be written as $H_{0}=\omega_{\mathrm{q}}S_{z}+\omega_{0}a^{\dagger}a$,
where $S_{z}=\sum_{j}\sigma_{j,z}$ accounts for the spin magnetic
moment. The quasi-particle states are therefore the resonant states
between the resonator photons and the qubit excitations, which diagonalize
the Hamiltonian $H=H_{0}+H_{\mathrm{int}}$ in the collective spin
formulation.

In order to make the interaction Hamiltonian block-diagonalizable
like that of Tavis-Cummings (TC) model, the set of operators $\{S_{\pm},S_{z}\}$
must conform to a SU(2) Lie algebra. Nonetheless, since $[S_{+},S_{-}]\neq2S_{z}$
because of the non-uniform coupling in the quasi-lattice, we approximate
the set as a deformed algebra by projecting the commutator into the
SU(2) group manifold through a trace over the Hilbert space spanned
by the quasi-lattice. Geometrically, this method associates with the
deformation of the Bloch sphere $\mathbb{S}^{2}$, the two-dimensional
projection from the 3-sphere $\mathbb{S}^{3}$ that is isomorphic
to SU(2), into an ellipsoid~\cite{ian12,ian14}. The degree of elongation
of the resulting ellipsoid from the original 2-sphere is captured
by the c-number deformation factor $\mathfrak{f}$ in the commutation
relation $[S_{+},S_{-}]=2\mathfrak{f}S_{z}$.

Algebraically, the deformation factor $\mathfrak{f}$ measures how
the new spin operators $\{S_{\pm},S_{z}\}$ differ from their counterparts
in a standard SU(2) algebra. In other words, as a function as $\ell$
and $N$, it figuratively quantifies the degree the quasi-lattice
resembles a normal lattice. Its value to the first-order approximation
under the Hilbert-Schmidt norm is $\mathfrak{f}=\sum_{j}g_{j}^{2}/N$.
With $g_{j}$ being the cosines defined above, we have

\begin{equation}
\mathfrak{f}=\frac{1}{2}+\frac{1}{4N}\left[1+\frac{\sin(2N-1)\pi\ell}{\sin\pi\ell}\right].\label{eq:deform_fac}
\end{equation}
When the qubits are inhomogeneously distributed and thus asynchronously
driven at uneven rates, $\ell$ takes a fractional value and $\mathfrak{f}$
adopts a positive sub-unit value. On the other hand, when (all qubits
fall on the antinodes on the resonator mode) or the limiting value
of zero (the inter-qubit spacing is negligible compared to the resonator
wavelength, as in a natural lattice),$\ell$ take an integer value
and $\mathfrak{f}$ reduces to one, signifying the quasi-lattice falls
back to a homogeneous lattice structure. 

After the projection-deformation process described above, the Hamiltonian
in Eq.~(\ref{eq:ham}) can be diagonalized in the usual angular momentum
states under TC-model, the eigenvalues of which read 
\begin{equation}
E_{u}=\omega_{\mathrm{q}}u+\varepsilon\label{eq:eig_enrg}
\end{equation}
for a quasi-particle ``excitation'' number $u=n+m$, which is distributed
among the photon count $n$ from the resonator and the spin moment
$m$ from the quasi-lattice. Since we have adopted the angular momentum
model, $u$ can take either integer or half-integer values greater
than $(-N/2)$. The second term $\varepsilon$ is the energy splitting
that accounts for the qubit-photon detuning $\Delta\omega=\omega_{0}-\omega_{\mathrm{q}}$
and the lattice-photon interaction with maximum coupling strength
$g$ introduced. At resonance $\Delta\omega=0$, $\varepsilon$ would
become a multiple of $g$ only. In brief, the resonator mode dresses
the quasi-lattice states into a series of clusters of states, grouped
by the value of $u$.

Given the eigenenergy in Eq.~(\ref{eq:eig_enrg}) for each $u$,
one can associate a time-independent Schroedinger equation with the
eigenvector 
\begin{equation}
\left|u,r\right\rangle =\sum_{n=0}^{u+r}c_{n}\left|n\right\rangle \otimes\left|r,m\right\rangle ,\label{eq:dressed_basis}
\end{equation}
which is parametrized additionally by the total spin number $r$ for
$N$ qubits. It has a $(u+r+1)$-fold degeneracy distributed among
different combinations of $n$ and $m$, which is split by the finite
interaction into finer levels with splitting $\varepsilon$, analogous
to ac-Stark shifts to an atom. Solving the time-independent eigen-equation
$H\left|u,r\right\rangle =E_{u}\left|u,r\right\rangle $ using Eq.~(\ref{eq:eig_enrg})-(\ref{eq:dressed_basis}),
one finds that both this splitting $\varepsilon$ and the coefficients
$c_{n}$ are determined \emph{a posteriori }by a recursive relation
\begin{equation}
c_{n}=\frac{\varepsilon-\Delta\omega(n-1)}{g\alpha_{r,m}\sqrt{n\mathfrak{f}}}c_{n-1}-\sqrt{\frac{n-1}{n}}\frac{\alpha_{r,m+1}}{\alpha_{r,m}}c_{n-2}\label{eq:coeff}
\end{equation}
with initial conditions $c_{u+r+1}=c_{-1}=0$. In the relation, $\alpha_{r,m}=\sqrt{(r-m)(r+m+1)}$
are the off-diagonal elements of the ladder operators. The initial
value $c_{0}$ can take arbitrary value since the subsequent coefficients
are all multiples of $c_{0}$, hence it can be regarded as a constant
arbitrarily chosen for normalization. The analytical expressions for
$c_{n}$ and $\varepsilon$ can be found by treating Eq.~(\ref{eq:coeff})
as a difference equation~\cite{ian12}. The coefficients $c_{n}$,
which represent the probability amplitudes of different mixed resonant
states between the quasi-lattice and the resonator photon, are not
only determined by the off-diagonal transition elements, but are also
determined by the deformation factor $\mathfrak{f}$. The latter is
thus a structure parameter of the quasi-lattice that decides the form
each photon is transferred to all the qubits, affecting how entanglement
is generated among the qubits.

Dynamically speaking, the quasi-particle states described by Eq.~(\ref{eq:dressed_basis})
are the stationary state of the lattice-cavity system when the qubits
are collectively resonant with the electromagnetic cavity mode. They
also serve as a set of transformed or dressed bases for the combined
system, which is parametrized by $u$ and $r$ after the transformation.
Conversely, any initial states given as a tensor product of an arbitrary
cavity Fock states $\sum_{n}d_{n}\left|n\right\rangle $ (e.g. $d_{n}$
could be $\alpha^{n}/\sqrt{n!}$ for a coherent state) and the lattice
angular momentum state $\left|r,m\right\rangle $ can be decomposed
in the dressed bases as
\begin{equation}
\left|\psi\right\rangle =\sum_{n}d_{n}\left|n\right\rangle \otimes\left|r,m\right\rangle =\sum_{u}\delta_{u}\left|u,r\right\rangle .\label{eq:trfm}
\end{equation}
Then, by regarding each sequence $\{c_{n}\}$, $\{d_{n}\}$, and $\{\delta_{n}\}$
as the column matrix $\mathbf{c}$, $\mathbf{d}$, and $\bm{\delta}$,
respectively, one can form the transformation matrix $[C]_{un}$ from
Eq.~(\ref{eq:dressed_basis}) to find $\bm{\delta}=C^{-1}\mathbf{d}$.

\section{Multi-photon resonances}

The coefficients of each resonant stationary state, as determined
by Eq.~(\ref{eq:coeff}), shows that the resonant state is unequally
contributed by a set of quasi-lattice configurations $\left|r,m\right\rangle $.
If the quasi-lattice is initialized to the ground configuration $\left|r,-r\right\rangle $,
then different numbers of photons are brought into resonance with
the quasi-lattice to form the resulting superposition state with different
$m$. In the case of superconducting quantum circuits, this is achieved
by sending pulse signals of different $\pi$ lengths into the qubits
whose level spacings are tuned in-resonance with the waveguide resonator~\cite{lucero12}.
In the following, a particular eigenstate $\left|u,r\right\rangle $
with a fixed $u$ is considered.

In a resonant process, the excitation number $u$ shared between the
photon Fock states and the quasi-lattice angular-momentum states is
conserved. The fixed $u$ permits both single-photon resonance (between
$\left|n=u+r\right\rangle $ and $\left|n=u+r-1\right\rangle $) and
multi-photon resonances (between $\left|n=u+r\right\rangle $ and
$\left|n=u-m\right\rangle $ where $m>-r+1$). In addition, the inhomogeneous
coupling quantified by the deformation factor $\mathfrak{f}$ plays
a part in Eq.~(\ref{eq:coeff}) for the multi-photon processes. To
investigate what effect it can exert, we look at the patterns of entanglements
formed on the quasi-lattice through its variation range and discern,
specifically, how these patterns associate with the two entanglement
types of $\left|\mathrm{GHZ}\right\rangle $ state and $\left|W\right\rangle $
state.

For a $N$-qubit quasi-lattice whence $r=N/2$, the $\left|\mathrm{GHZ}\right\rangle $
state is generated from an $N$-photon resonance between the lattice
states $\left|r,-r\right\rangle $ and $\left|r,r\right\rangle $.
For a quasi-particle excitation with excitation number $u$, it constitutes
the two terms $c_{N}\left|N\right\rangle \otimes\left|r,-r\right\rangle +c_{0}\left|0\right\rangle \otimes\left|r,r\right\rangle $
within the decomposition of $\left|u,r\right\rangle $ for any $u$.
Whereas, the $\left|W\right\rangle $ state is generated from an one-photon
resonance, i.e. it associates with the terms $c_{u+r}\left|u+r\right\rangle \otimes\left|r,-r\right\rangle +c_{u+r-1}\left|u+r-1\right\rangle \otimes\left|r,-r+1\right\rangle $,
depending on a particular $u$ for the total system excitation. (If
the inverted $\left|\bar{W}\right\rangle $ state is also considered,
the term $c_{u-r+1}\left|r,r-1\right\rangle $ should be included.)
The general consideration is therefore whether the distribution of
the coefficients $\{c_{n}\}$ as determined by Eq.~(\ref{eq:coeff})
would concentrate on $c_{0}$ and $c_{u-r+1}$ (and $c_{u-r+1}$)
if the deformation factor $\mathfrak{f}$ takes value away from either
0 or 1, indicating an inhomogeneous coupling and thus asynchronous
excitation scheme.

\begin{figure}
\includegraphics[clip,width=8cm]{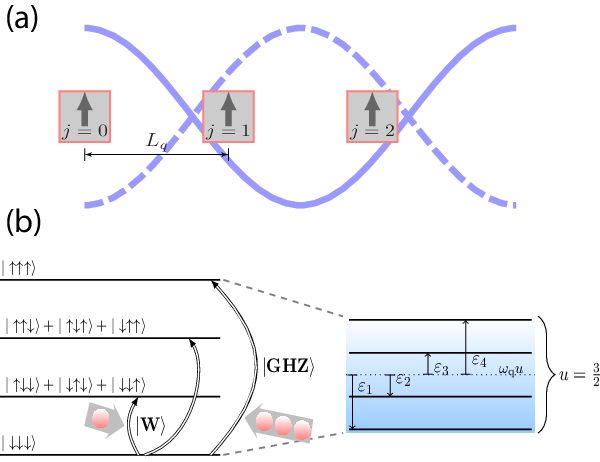}

\caption{(color online) (a) Illustration of a 3-qubit quasi-lattice with $\ell=2/3$.
(b) The level diagram of the 3-qubit quasi-lattice, with the ground
state resonant with multiple photons, forming two classes of entanglement
states ($\left|\mathrm{GHZ}\right\rangle $ and $\left|W\right\rangle $
states). The energy levels are effectively dressed into a cluster
of states of the same ``excitation'' number $3/2$, split by shifts
$\varepsilon_{1,2,3,4}$ due to the lattice-photon interaction.~\label{fig:3-qubit}}

\end{figure}

Since solving Eq.~(\ref{eq:coeff}) for an expression of $c_{n}$
would not illustrate clearly the change in the distribution and, more
importantly, researches on general $N$-qubit entanglement measures
are plural with no particular established ones, we examine a simplified
case of a 3-qubit quasi-lattice for further considerations. The 3-qubit
system has two metrics of entanglement that can distinctively discriminate
a $\left|\mathrm{GHZ}\right\rangle $ state from a $\left|W\right\rangle $
state. The model is illustrated in Fig.~\ref{fig:3-qubit}(a) and
the resonant processes are illustrated in Fig.~\ref{fig:3-qubit}(b)
for the case of $u=\frac{3}{2}$, for which the clustered dressed
level is split into four sublevels for the four values of the splitting
$\varepsilon$. The quasi-lattice at ground state $\left|\frac{3}{2},-\frac{3}{2}\right\rangle $
can be resonant with both $\left|\frac{3}{2},\frac{3}{2}\right\rangle $
through a triple-photon process to generate the $\left|\mathrm{GHZ}\right\rangle $
state and $\left|\frac{3}{2},-\frac{1}{2}\right\rangle $ through
a single-photon process to generate the $\left|W\right\rangle $ state,
as shown on the left side of Fig.~\ref{fig:3-qubit}(a). The state
$\left|\frac{3}{2},\frac{1}{2}\right\rangle $ is the inverted $\left|\bar{W}\right\rangle $
state, which is equivalent to $\left|\frac{3}{2},-\frac{1}{2}\right\rangle $
up to an LOCC operation and can be generated from the quasi-lattice
ground state through a double-photon resonance.

To see how asynchronous excitations benefit the generation of $\left|\mathrm{GHZ}\right\rangle $
type of entanglement in a stationary state, we calculate two coefficients
$c_{0}$ and $c_{3}$ for $\left|u=\frac{3}{2}\right\rangle $, which
associate respectively with the eigenstates $\left|\uparrow\uparrow\uparrow\right\rangle $
and $\left|\downarrow\downarrow\downarrow\right\rangle $ of the $\left|\mathrm{GHZ}\right\rangle $
state. The analytical solution to the two coefficients is subject
to a common normalizing factor and their ratio can be compactly written
as 
\begin{equation}
\rho=\frac{c_{0}}{c_{3}}=\frac{6\sqrt{6}g^{3}\mathfrak{f}^{3/2}}{\varepsilon(\varepsilon-\Delta\omega)(\varepsilon-2\Delta\omega)-(11\varepsilon-6\Delta\omega)g^{2}\mathfrak{f}}\label{eq:ratio}
\end{equation}
By taking the derivative $d\rho/d\mathfrak{f}$, one finds the extrema
of the ratio at
\begin{equation}
\mathfrak{f}_{\ast}=\frac{3\varepsilon(\varepsilon-\Delta\omega)(\varepsilon-2\Delta\omega)}{11\varepsilon-6\Delta\omega}\label{eq:f_max}
\end{equation}
which is neither 0 nor 1. Checking further that $d^{2}\rho/d\mathfrak{f}^{2}|_{\mathfrak{f}_{\ast}}<0$,
one finds $c_{0}$ would increase under inhomogeneous coupling, favoring
$\left|\mathrm{GHZ}\right\rangle $ state generation. Substituting
Eq.~(\ref{eq:f_max}) into Eq.~(\ref{eq:deform_fac}), one has the
equation $U_{4}(\cos\pi\ell)-12\mathfrak{f}_{\ast}+7=0$, where $U_{4}$
is the fourth-order Chebyshev polynomial of the second kind. Since
the polynomial has four zeros, ignoring the two zeros on the negative
axis of the even function $\cos\pi\ell$ shows that the ratio $\rho$
admits two maximizing values of $\ell$.

\begin{figure}
\includegraphics[bb=50bp 180bp 560bp 610bp,clip,width=8.5cm]{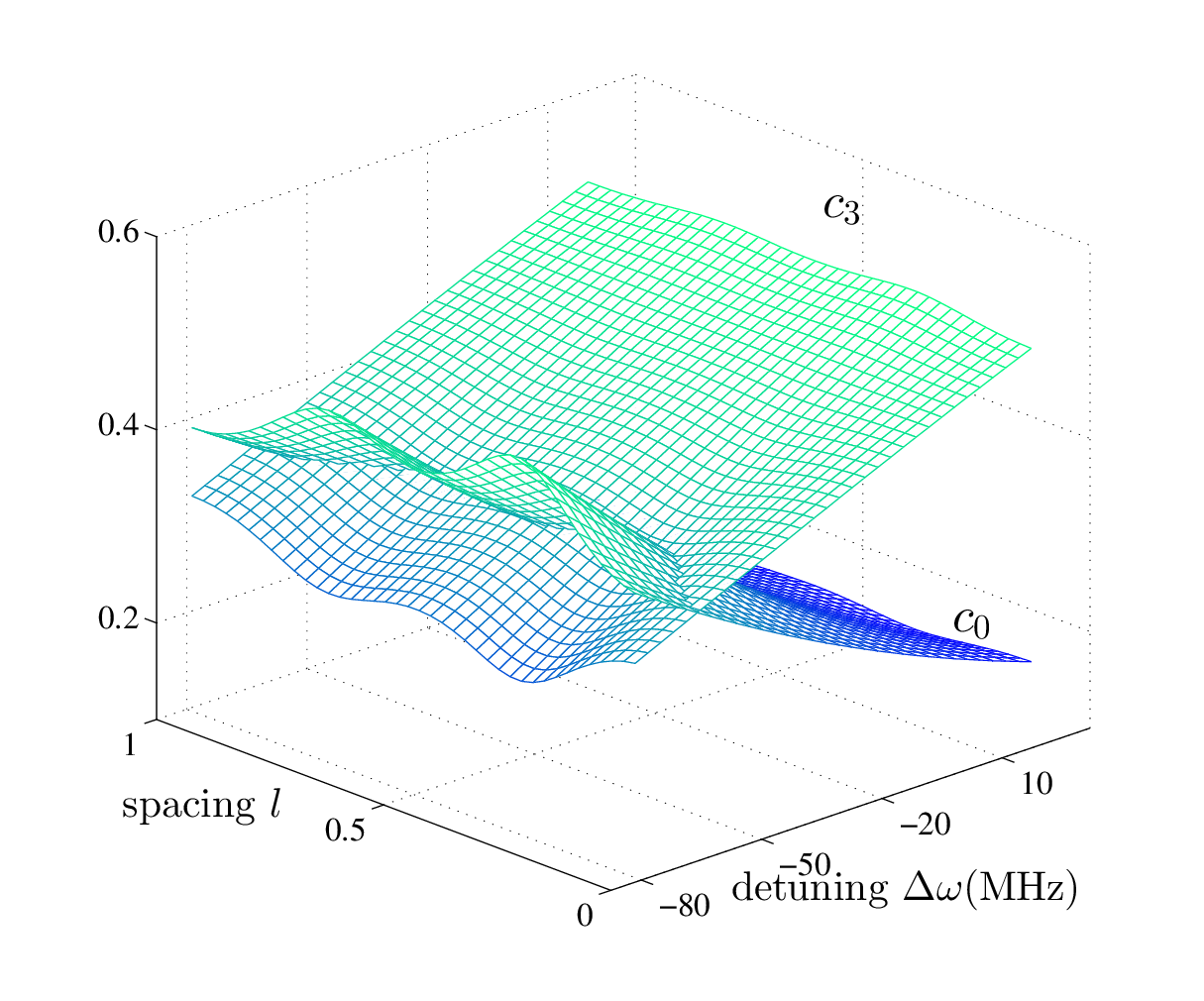}

\caption{(color online) The distribution of the normalized coefficients $c_{0}$
of the ground state and $c_{3}$ of the excited state versus the relative
qubit spacing and the detuning for the clustered state $\left|u=\frac{3}{2}\right\rangle $.~\label{fig:coefficients}}

\end{figure}

Using the experimental data given in Ref.~\cite{fink09}, the numerical
solutions of $c_{0}$ and $c_{3}$ are plotted for the highest split
state of $\varepsilon$ in Fig.~\ref{fig:coefficients} against the
relative spacing $\ell$ on one axis and against the detuning $\Delta\omega$
on the other. We observe that at exact lattice-photon resonance or
at positive detuning, the coefficient $c_{0}$ (for the all spin-up
state) is close to zero, meaning the probability of triple-photon
resonance is low. In the range of negative detuning, $c_{0}$ starts
to increase while the $c_{3}$ (for the all spin-down state) correspondingly
decreases and their variations over the relative qubit spacing $\ell$
intensify. The plot shows that $c_{0}$ obtains maximum ($c_{3}$
correspondingly obtains minimum) at two values: $\ell\approx0.3$
and $\ell\approx0.7$, verifying our analytical predictions above.
Being neither 0 nor 1 and symmetric about the middle point $\ell=0.5$,
the maxima show that inhomogeneous coupling increases the probability
of multi-photon absorptions that lead to the generation of $\left|\mathrm{GHZ}\right\rangle $
state.

\section{Multipartite entanglement }

To further verify the results, we put the computed coefficients into
two metrics of entanglement measures. One is the tripartite concurrence
introduced by Carvalho \emph{et al.}~\cite{carvalho04} that generalizes
the original bipartite concurrence given by Wootters~\cite{wootters98}.
The other is the 3-tangle introduced by Coffman \emph{et al.}~\cite{coffman00}.
Whereas 3-tangle quantifies only the three-party total non-separability,
tripartite concurrence accounts for both partial and total non-separabilities.
In other words, when accounting the distributed entanglement among
a 3-qubit quasi-lattice, tripartite concurrence does not distinguish
the totally non-separable $\left|\mathrm{GHZ}\right\rangle $ state
and the partially non-separable $\left|W\right\rangle $ state. For
example, the state $\left|\uparrow\downarrow\downarrow\right\rangle +\left|\downarrow\uparrow\downarrow\right\rangle =\left(\left|\uparrow\downarrow\right\rangle +\left|\downarrow\uparrow\right\rangle \right)\otimes\left|\downarrow\right\rangle $
is partially non-separable and not totally non-separable. Hence, tripartite
concurrence quantifies both types of non-separability as entanglement
on an equal footing. 3-tangle, in contrast, is designed specifically
to neglect the partial entanglement; it only measures the so-called
residual entanglement that is left in a system after the partial entanglement
are removed, i.e. it is non-zero only for $\left|\mathrm{GHZ}\right\rangle $
state~\cite{dur00,lohmayer06}. Here, by contrasting the two metrics
over the same range of variation of a parameter, one is able to distinguish
the indications of $\left|\mathrm{GHZ}\right\rangle $ state and $\left|W\right\rangle $
state in a quantum system.

\begin{figure}
\includegraphics[bb=12bp 40bp 530bp 405bp,clip,width=7.8cm]{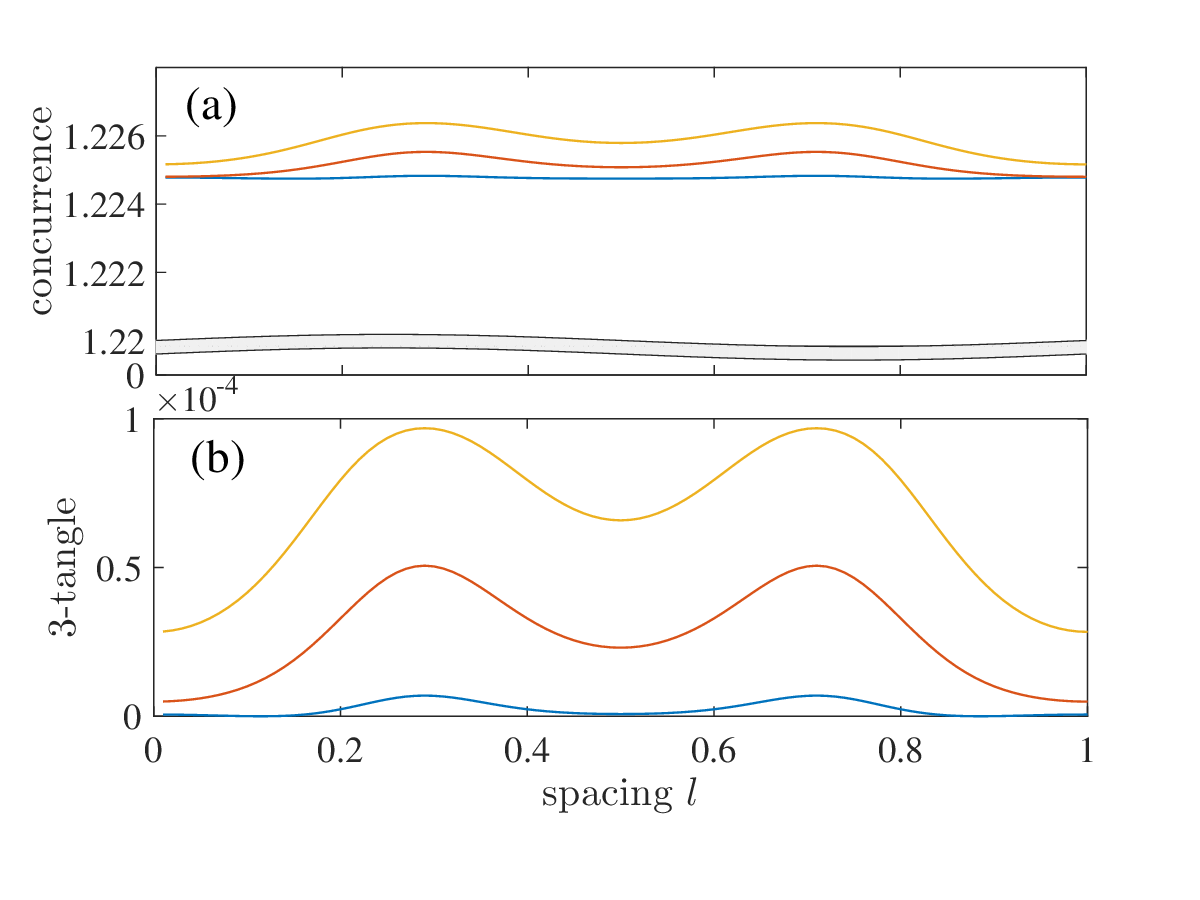}

\caption{(color online) Tripartite entanglements measured in two metrics: (a)
the tripartite concurrence and (b) the 3-tangle, at three values of
detuning $\Delta\omega=-66$MHz (yellow), $\Delta\omega=-92$MHz (orange),
and $\Delta\omega=-117$Mhz (blue). In (a), the variation of the concurrence
over $\ell$ is small compared to its average. We break the vertical
axis to magnify the scope of variation.~\label{fig:entanglement}}
\end{figure}

Figure~\ref{fig:entanglement}(a) plots the tripartite concurrence
as a function of $\ell$ over the range $(0,1)$ for three detuning
values of $\Delta\omega$. Figure~\ref{fig:entanglement}(b) plots
3-tangle against $\ell$ for the same three detunings. Other parameters
are set to the same values of those used in Fig.~\ref{fig:coefficients}.
For the three values of detunings, the concurrences are all close
to the maximum attainable value of $1.25$, showing that the resonance
between the quasi-lattice and the resonator photon generates highly
entangled states among the qubits. It is particularly noticeable that
the concurrence peaks at the same values of $\ell$ as those for the
coefficient $c_{0}$, showing that multi-photon absorptions facilitated
by inhomogeneous coupling also assist entanglement generation. However,
the assistance is not large. For example, at $\Delta\omega=-92$MHz,
the increase at the $\ell=0.3$ peak from $\ell=0$ is about 0.06\%.

3-tangle demonstrate similar trends in variation over $\ell$ and
peaks at the same points, even though its mean retains a small value
throughout. Nevertheless, the variation is much more apparent: at
$\Delta\omega=-92$MHz, 3-tangle is close to zero when the quasi-lattice
is homogeneous at the two ends of abscissa; whereas at the $\ell=0.3$
peak, the increase over the two ends is about 94-fold.

These variations coincide with our perceptions. First, multi-photon
processes are probabilistically rarer than single-photon processes,
prompting lesser generations of $\left|\mathrm{GHZ}\right\rangle $
state than $\left|W\right\rangle $ state in general and giving a
much lower mean in 3-tangle than in tripartite concurrence. Second,
as predicted from our analysis of the coefficients, inhomogeneous
coupling exemplified by the peaking values at $\ell\approx0.3$ and
$\ell\approx0.7$ facilitates entanglement in general. If one regards
concurrence and 3-tangle commensurate, then the contribution of 3-tangle
to the concurrence increase at its maximal points occupies only about
five to seven percent, showing not just $\left|\mathrm{GHZ}\right\rangle $
state generation but overall entangled state generation is facilitated.
Third, inhomogeneous coupling does not assist but rather impedes single-photon
processes because inhomogeneity decreases the qubit-photon coupling
strength on average from the maximal value $g$. Since the coupling
strength corresponds to the transition rate of each qubit, the probability
of exciting any one qubit among all in the quasi-lattice from $\left|\frac{3}{2},-\frac{3}{2}\right\rangle $
to $\left|\frac{3}{2},-\frac{1}{2}\right\rangle $ is lowered. Our
computation verifies that the coefficient $c_{2}$ for $\left|\frac{3}{2},-\frac{1}{2}\right\rangle $
indeed dips at $\ell=0.3$ and $0.7$. Nevertheless, inhomogeneous
coupling does facilitate double-photon and triple-photon resonances,
increasing the entanglement for inverted$\left|\bar{W}\right\rangle $
state and $\left|\mathrm{GHZ}\right\rangle $ state, respectively,
especially the latter. Tripartite concurrence, accounting for all
types of entanglement, is consequently compensated and adversely peaks
at these originally dipping positions, resulting in a rather flat
shape on average.

\section{Conclusions and discussions}

We have demonstrated the role of inhomogeneous coupling in elevating
the generation of $\left|\mathrm{GHZ}\right\rangle $ type of entanglement
in a quasi-lattice of qubits, The analysis conducted above is confined
to stationary states that form the bases of the dressed lattice-resonator
system and it requires the transformation in Eq.~(\ref{eq:trfm})
to compute the entanglement for an arbitrary state $\left|\psi\right\rangle $.
How the entanglement would approach the stationary measures along
with the evolution of $\left|\psi(t)\right\rangle $ however remains
unknown and require future studies. But, in the least, from our studies
of resonator-mediated double-cavity systems~\cite{huan15} and currently
of cavity-coupled two-qubit system, we expect that: (i) the entanglement
between the resonator and any one qubit it directly couples to can
be transferred to that between two indirectly coupled qubits; (ii)
the Rabi oscillations of the qubits render the stationary measures
computed here an envelop in the time domain within which the entanglement
would oscillate; (iii) it requires a finite time for the entanglement
envelop to approach a stationary value, analogous to how $N$ atoms
would radiate after a finite delay in a superradiant state.

\acknowledgments
The author thanks X.~G.~Wang, Q.~Ai, S.~Ashhab, and E.~Lucero
for stimulating discussion. The research is supported by FDCT of Macau
under Grant 013/2013/A1, University of Macau under Grants MRG022/IH/2013/FST
and MYRG2014-00052-FST, and the National Natural Science Foundation
of China under Grant No.~11404415.

\end{document}